\documentclass[final,5p,times,twocolumn]{elsarticle}
 \usepackage{graphics}
 \usepackage{array}
\usepackage{amsmath}
 \usepackage{amsthm}

 \usepackage{lineno}
 \biboptions{comma,square,compress}

\begin{document}
\begin{frontmatter}



\title{Advantages of the multinucleon transfer reactions based on $^{238}$U target for producing neutron-rich isotopes around $N=126$ }

\author[1]{Long Zhu \corref{cor}}
\author[2]{Cheng Li}
\author[1]{Jun Su}
\author[1]{Chen-Chen Guo}
\author[1]{Wei Hua}
\address[1]{Sino-French Institute of Nuclear Engineering and Technology, Sun Yat-sen University, Zhuhai 519082, China}
\address[2]{Beijing Radiation Center, Beijing 100875, China}
\cortext[cor]{zhulong@mail.sysu.edu.cn}

\begin{abstract}
The mechanism of multinucleon transfer (MNT) reactions for producing neutron-rich heavy nuclei around $N=126$ is investigated within two different theoretical frameworks: dinuclear system (DNS) model and isospin-dependent quantum molecular dynamics (IQMD) model. The effects of mass asymmetry relaxation, $N/Z$ equilibration, and shell closures on production cross sections of neutron-rich heavy nuclei are investigated. For the first time, the advantages for producing neutron-rich heavy nuclei around $N=126$ are found in MNT reactions based on $^{238}$U target. We propose the MNT reactions with $^{238}$U target for producing unknown neutron-rich heavy nuclei around $N=126$ in the future.
\end{abstract}


\end{frontmatter}

% \linenumbers

%\section{}
%\label{}\
\section{\label{int}Introduction}
The neutron-rich heavy nuclei (NRHN) around $N=126$ are not only interesting in nuclear structure, but also contribute significantly to understanding the mechanism of heavy elements synthesis in r-process. As a consequence, much effort has been made for producing NRHN around $N=126$ in recent years \cite{Corradi01,Zagrebaev05,Barrett01,Kozulin02,Zhu04,Watanabe01}. Several reactions, such as $^{136}$Xe + $^{208}$Pb \cite{Zagrebaev05,Barrett01,Kozulin02} and $^{136}$Xe + $^{198}$Pt \cite{Watanabe01,Li02}, are proposed to produce NRHN in this region. Huge advantage of production cross section is noticed in multinucleon transfer (MNT) process in comparison to the approach of projectile fragmentation \cite{Watanabe01}. However, no new isotopes with $N=126$ is observed directly in recent experiments. Further developments of experimental detection and separation capabilities are needed. Also, it is desirable to investigate the MNT mechanism theoretically, which can give more clues to find favorable combinations to produce unknown nuclei.

Strong dependence of the nucleon flow on the shell effects in MNT reactions around Coulomb barrier was found both theoretically \cite{Zagrebaev01,Zhu09} and experimentally \cite{Kozulin03}. Due to influence of shell closures $N=82$ and $N=126$, the enhancement of yield in transtarget region was noticed in the reaction $^{160}$Gd + $^{186}$W. The behavior that the nucleons are transferred from the lighter partner to the heavy one is called inverse quasifission (QF) process, which could be one candidate for producing NRHN around $N=126$. In Refs. \cite{Zhu04,Iwata01,Zhu05,Adamian02}, it was found that due to $N/Z$ equilibration, the projectiles with large neutron excesses show great advantage of production cross sections of neutron-enriched transtarget nuclei. For radioactive beams, although the neutron excesses are large, the beam intensities are usually lower than stable ones. Also, it is very difficult to detect the target-like fragments (TLF) in MNT reactions with present experimental equipments. The MNT reactions with $^{238}$U projectile and lighter targets had been performed many years ago by Mayer $\emph{et al}$ \cite{Mayer01}. The structure effects on nucleon flow were observed. Also, theoretically, protons transfer from $^{238}$U to lighter partner was explained by enhanced neck evolution \cite{Sekizawa04}. Nevertheless, the systematic study on the advantages of MNT reactions with $^{238}$U for producing NRHN around $N=126$ has not yet been given.

In this work, in order to optimize the reaction combinations for producing unknown NRHN around $N=126$, based on the dinuclear system (DNS) model and isospin-dependent quantum molecular dynamics (IQMD) model, the reactions $^{176}$Yb + $^{238}$U, $^{186}$W + $^{238}$U, $^{192}$Os + $^{238}$U, $^{198}$Pt + $^{238}$U, $^{176}$Yb + $^{170}$Er, $^{186}$W + $^{160}$Gd, $^{192}$Os + $^{154}$Sm, and  $^{198}$Pt + $^{136}$Xe to produce unknown NRHN are investigated at incident energies of $E_{\textrm{c.m.}}=$ 621, 660, 684, 711, 470, 475, 478, and 458 MeV, respectively, which are 1.15 times of Coulomb barriers. The DNS model in combination with the GEMINI code \cite{Zhu04,Zhu05,Zhu01,Wen01} and QMD type models \cite{Zhu03,Zhao01,Wang01,Feng01,Li01} have been developed and successfully used in investigation of nuclear reactions around Coulomb barrier, including multinucleon transfer reactions. Two approaches show different points of view on multinucleon transfer process. We expect that the reactions with $^{238}$U target would show great advantages for producing NRHN around $N=126$ based on three conjectures. (i) The mass asymmetry relaxation would promote the nucleons transferring from $^{238}$U to light partners; (ii) The neutron closed shell $N=126$ could attract the neutrons flow from $^{238}$U to light partners; (iii) The $^{238}$U shows large value of $N/Z$ ratio and could enhance the neutron-richness of projectile-like products. Also, with consideration of direct kinematics, the reactions with heavier target $^{238}$U could be easier performed, although the inverse kinematics reactions, such as $^{238}$U + $^{110}$Pd, had been performed experimentally \cite{Mayer01}.

The article is organized as follows. In Sec. \ref{model}, we briefly describe the theoretical models. The results and discussion are presented in Sec. \ref{result}. Finally, we summarize the main results in Sec. \ref{summary}.

\section{\label{model}Models}
The DNS+GEMINI model was improved by consideration of deformation degree of freedom and the temperature dependence of shell correction \cite{Zhu09,Zhu01}. In order to decrease the number of collective parameters, one unified dynamical deformation $\beta_{2}$ is used instead of two independent $\delta \beta_{2}^{1}$ and $\delta \beta_{2}^{2}$. $C_{1}\delta \beta_{2}^{1}=C_{2}\delta \beta_{2}^{2}$; $\delta \beta_{2}^{1}+ \delta \beta_{2}^{2}=2\beta_{2}$ \cite{Zagrebaev01}. $\beta_{2}^{1}=\beta_{2}^{\textrm{p}}+\delta \beta_{2}^{1}$ and $\beta_{2}^{2}=\beta_{2}^{\textrm{t}}+\delta \beta_{2}^{2}$ are quadrupole deformation parameters of projectile-like fragments (PLF) and TLF, respectively. $\beta_{2}^{\textrm{p}}$ and $\beta_{2}^{\textrm{t}}$ are static deformation parameters of projectile and target, respectively, which are taken from Ref. \cite{Moller01}. $C_{1,2}$ are the LDM stiffness parameters of the fragments, the description of which can be seen in Ref. \cite{Myers01}.

The master equation can be written as
\begin{flalign}
\begin{split}\label{master}
&\frac{dP(Z_{1},N_{1},\beta_{2},t)}{dt}\\
&=\sum_{Z_{1}^{'}}W_{Z_{1},N_{1},\beta_{2};Z_{1}^{'},N_{1},\beta_{2}}(t)[d_{Z_{1},N_{1},\beta_{2}}P(Z_{1}^{'},N_{1},\beta_{2},t)\\
&-d_{Z_{1}^{'},N_{1},\beta_{2}}P(Z_{1},N_{1},\beta_{2},t)]\\
&+\sum_{N_{1}^{'}}W_{Z_{1},N_{1},\beta_{2};Z_{1},N_{1}^{'},\beta_{2}}(t)[d_{Z_{1},N_{1},\beta_{2}}P(Z_{1},N_{1}^{'},\beta_{2},t)\\
&-d_{Z_{1},N_{1}^{'},\beta_{2}}P(Z_{1},N_{1},\beta_{2},t)]\\
&+\sum_{\beta_{2}^{'}}W_{Z_{1},N_{1},\beta_{2};Z_{1},N_{1},\beta_{2}^{'}}(t)[d_{Z_{1},N_{1},\beta_{2}}P(Z_{1},N_{1},\beta_{2}^{'},t)\\
&-d_{Z_{1},N_{1},\beta_{2}^{'}}P(Z_{1},N_{1},\beta_{2},t)],
\end{split}
\end{flalign}
where $P(Z_{1},N_{1},\beta_{2},t)$ is the distribution probability for the fragment 1 with proton number $Z_{1}$ and neutron number $N_{1}$ at time $t$. $P(Z_{1},N_{1},\beta_{2},t)$ is actually equivalent to $P(\eta_{Z},\eta_{A},\beta_{2},t)$. $\eta_{Z}$ and $\eta_{A}$ are charge and mass asymmetries, respectively. $Z_{1}$, $N_{1}$, and $\beta_{2}$ in the master equation are collective variables. $\beta_{2}$ influences the potential energy surface, and then affects the distribution probabilities of $Z_{1}$ and $N_{1}$. $W_{Z_{1},N_{1},\beta_{2};Z_{1}^{'},N_{1},\beta_{2}}$ denotes the mean transition probability from the channel ($Z_{1}$, $N_{1}$, $\beta_{2}$) to ($Z_{1}^{'}$, $N_{1}$, $\beta_{2}$), which is similar to $N_{1}$ and $\beta_{2}$.
 $d_{Z_{1},N_{1},\beta_{2}}$ is
the microscopic dimension (the number of channels) corresponding to the macroscopic state ($Z_{1}$, $N_{1}$, $\beta_{2}$) \cite{Norenberg01}. For the degrees of freedom of charge and neutron number, the sum is taken over all possible proton and neutron numbers that fragment 1 may take, but only one nucleon transfer is considered in the model ($Z_{1}^{'}=Z_{1}\pm1$; $N_{1}^{'}=N_{1}\pm1$). For the $\beta_{2}$, we take the range of -0.5 to 0.5. The evolution step length is 0.01. $\beta_{2}^{'}=\beta_{2}\pm0.01$. The transition probability is related to the local excitation energy \cite{Zhu01,Ayik01}.

The potential energy surface (PES) is defined as
\begin{flalign}
\begin{split}
 U(Z_{1}, N_{1}, \beta_{2}, R_{\textrm{cont}})=&\Delta(Z_{1}, N_{1})+\Delta(Z_{2}, N_{2})\\
 &+V_{\textrm{cont}}(Z_{1}, N_{1}, \beta_{2}, R_{\textrm{cont}}).
\end{split}
\end{flalign}
Here, $\Delta(Z_{i}, N_{i})$ ($i=1$, 2) is mass excess of the fragment $i$, including the paring and shell corrections as shown in Ref. \cite{Zhu09}.

\begin{figure}
\begin{center}
\includegraphics[width=8.8cm,angle=0]{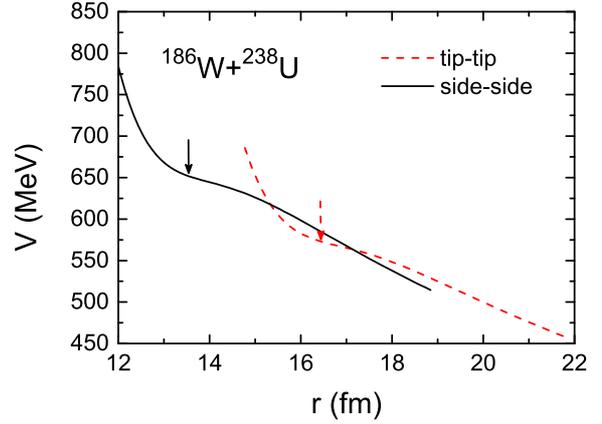}
\caption{\label{poten} (Color online.) Interaction potentials in entrance channel for the reaction $^{186}$W + $^{238}$U with different collision orientations. The arrows show the contact positions in entrance channel.}
\end{center}
\end{figure}

The effective nucleus-nucleus interaction potential $V_{\textrm{cont}}(Z_{1}, N_{1},\beta_{2}, R_{\textrm{cont}})$ between fragments 1 and 2 can be written as
\begin{flalign}
\begin{split}
 V_{\textrm{cont}}(Z_{1}, N_{1},\beta_{2}, R_{\textrm{cont}})&=V_{\textrm{N}}(Z_{1}, N_{1},\beta_{2}, R_{\textrm{cont}})+\\
V_{\textrm{C}}(Z_{1}, N_{1},\beta_{2}, R_{\textrm{cont}})+
&\frac{1}{2}C_{1}(\beta_{2}^{1}-\beta_{2}^{\textrm{p}})^{2}+\frac{1}{2}C_{2}(\beta_{2}^{2}-\beta_{2}^{\textrm{t}})^{2}.
\end{split}
\end{flalign}
Here, for the reactions with no potential pockets, the position where the nucleon transfer process takes place can be obtained with the equation: $R_{\textrm{cont}}=R_{1}(1+\beta_{2}^{1}Y_{20}(\theta_{1}))+R_{2}(1+\beta_{2}^{2}Y_{20}(\theta_{2}))+0.7$ fm. Here, $R_{1,2}=1.16A_{1,2}^{1/3}$. $\theta_{1}=\theta_{2}=0.$ The detailed description of nuclear potential and Coulomb potential can be seen in Refs. \cite{Zhu09,Bao01,Wong01}. Figure \ref{poten} shows the interaction potentials in entrance channel in the reaction $^{186}$W + $^{238}$U with different collision orientations. The arrows show the contact positions. It is shown that the contact positions are near the relatively flat parts of interaction potential curves. For side-side orientation, the entrance channel interaction potential is very high, which results in very short interaction time. Therefore, the production yields of the primary exotic nuclei will be lower than those in tip-tip collisions. For the reaction with incident energy around the Coulomb barrier, due to strong repulsive force, it is usually hard to form a DNS in side-side collision orientation. Usually, the PES is calculated in tip-tip orientation. In our code, the deformation degree of freedom is included self-consistently.

The local excitation energy of the DNS is determined by
\begin{flalign}
\begin{split}
E^{*}_{\textrm{DNS}}(Z_{1},N_{1},\beta_{2},J,t)=&E_{\textrm{diss}}(J,t)-
[U(Z_{1}, N_{1}, \beta_{2}, R_{\textrm{cont}})\\
&-U(Z_{\textrm{p}}, N_{\textrm{p}},\beta_{2}, R_{\textrm{cont}})].
\end{split}
\end{flalign}
Here,
\begin{flalign}
\begin{split}
E_{\textrm{diss}}(J,t)=&E_{\textrm{c.m.}}-V_{\textrm{cont}}(Z_{\textrm{p}}, N_{\textrm{p}},\beta_{2}, R_{\textrm{cont}})\\
&-\frac{(J^{'}(t)\hbar)^{2}}{2\zeta_{\textrm{rel}}}-E_{\textrm{rad}}(J,t),
\end{split}
\end{flalign}
where, $J^{'}(t)$ ($=J_{\textrm{st}}+(J-J_{\textrm{st}})e^{-t/\tau_{J}}$) is the relative angular momentum at time $t$. $J$ is initial entrance angular momentum.  $J_{\textrm{st}}=\frac{\zeta_{\textrm{rel}}}{\zeta_{\textrm{tot}}}J$. $\zeta_{\textrm{rel}}$ and $\zeta_{\textrm{tot}}$ are the relative and total moments of inertia, respectively. $\tau_{J}=12\times10^{-22}$ s.
$E_{\textrm{rad}}(J,t)=[E_{\textrm{c.m.}}-V_{\textrm{cont}}(Z_{\textrm{p}}, N_{\textrm{p}},\beta_{2}, R_{\textrm{cont}})-\frac{(J\hbar)^{2}}{2\zeta_{\textrm{rel}}}]e^{-t/\tau_{R}}$. $\tau_{R}$ ($=2\times10^{-22}$ s) is the characteristic relaxation time of radial energy.

The production cross sections of the primary products in transfer reactions can be calculated as follows:
\begin{flalign}
\begin{split}
 \sigma_{\textrm{pr}}(Z_{1},N_{1})=
 \frac{\pi\hbar^{2}}{2\mu E_{\textrm{c.m.}}}\sum_{J=0}^{J_{\textrm{max}}}(2J+1)[T_{\textrm{cap}}
 \sum_{\beta_{2}} P(Z_{1},N_{1},\beta_{2},\tau_{\textrm{int}})].
\end{split}
\end{flalign}
Here, the second sum is taken over all possible $\beta_{2}$ that may take. The interaction time $\tau_{\textrm{int}}$, which is strongly affected by entrance angular momentum, is calculated with the method shown in Refs. \cite{Li04,Wolschin01,Zhu09}. $T_{\textrm{cap}}$ is the capture probability. Usually, capture probability depends on the entrance angular momentum \cite{Adamian01,Mandaglio01}. Because there are no potential pockets for the reactions in this work (there are no ordinary barriers: the potential energies of these nuclei are everywhere repulsive) and the incident energies are above the interaction potentials at the contact configurations for different entrance angular momentum, the $T_{\textrm{cap}}$ is estimated as 1 \cite{Penion01}.

The statistical model GEMINI \cite{Charity01} is used to treat the sequential statistical evaporation of excited fragments. Assuming the situation of thermal equilibrium, the sharing of the excitation energy between the primary fragments is assumed to be proportional to their masses. Subsequent de-excitation cascades of the excited fragments via emission of light particles (neutron, proton, and $\alpha$) and $\gamma$ rays competing with the fission process are taken into account, which lead to the final mass distribution of the reaction products. The excitation energy of system depends on the entrance angular momentum. Therefore, the de-excitation of fragments in different angular momentums is taken into account.

In the IQMD model, the effective interaction potential energy is written as the sum of Coulomb interaction potential energy $U_{Coul}$ and the nuclear interaction potential energy $U_{loc}=\int V_{loc} (\textbf{r})d\textbf{r}$. $V_{loc}$ is potential energy density that is obtained from the effective Skyrme interaction
\begin{flalign}
\begin{split}\label{sky}
V_{loc}=&\frac{\alpha}{2}\frac{\rho^{2}}{\rho_{0}}+\frac{\beta}{\gamma+1}\frac{\rho^{\gamma+1}}{\rho_{0}^{\gamma}}+\frac{g_{sur}}{2\rho_{0}}(\nabla \rho)^{2}+g_{\tau}\frac{\rho^{\eta+1}}{\rho^{\eta}_{0}}\\
&+\frac{C_{s}}{2\rho_{0}}[\rho^{2}-k_{s}(\nabla \rho)^{2}]\delta^{2}.
\end{split}
\end{flalign}
Here, $\delta$ is the isospin asymmetry. The parameters are shown in Table \ref{tab:table1}.
\begin{table}
\small
\caption{\label{tab:table1} The parameter sets in Eq. (\ref{sky})}
\begin{center}
\setlength{\tabcolsep}{1.2mm}
\begin{tabular}{ccccccccc}
      \hline
      \hline
                 $\alpha$ & $\beta$ & $\gamma$  & $g_{sur}$ & $g_{\tau}$ & $\eta$ & $C_{s}$ & $\kappa_{s}$ & $\rho_{0}$\\

                 $MeV$  &  $MeV$   &  & $MeVfm^{2}$  & $MeV$ &  & $MeV$ & $fm^{2}$ & $fm^{-3}$\\
      \hline
                 -207  &  138  &  7/6 &  16.5  &  13
                 &5/3& 34 & 0.08 & 0.165 \\
      \hline
      \hline
\end{tabular}
\end{center}
\end{table}
The density distribution in the coordinate space is given by
\begin{equation}
\rho(r,t)=\sum\limits_{i}\frac{1}{(2\pi
L)^{3/2}}\exp[-\frac{(\textbf{r}-\textbf{r}_{i}(t))^{2}}{2L}],
\end{equation}
where, $L$ [$=(0.09A^{1/3}+0.88$)$^{2}$ fm$^{2}$] is the square of the Gaussian wave pocket width.

\section{\label{result}Results and discussion}

\begin{figure}
\begin{center}
\includegraphics[width=8.8cm,angle=0]{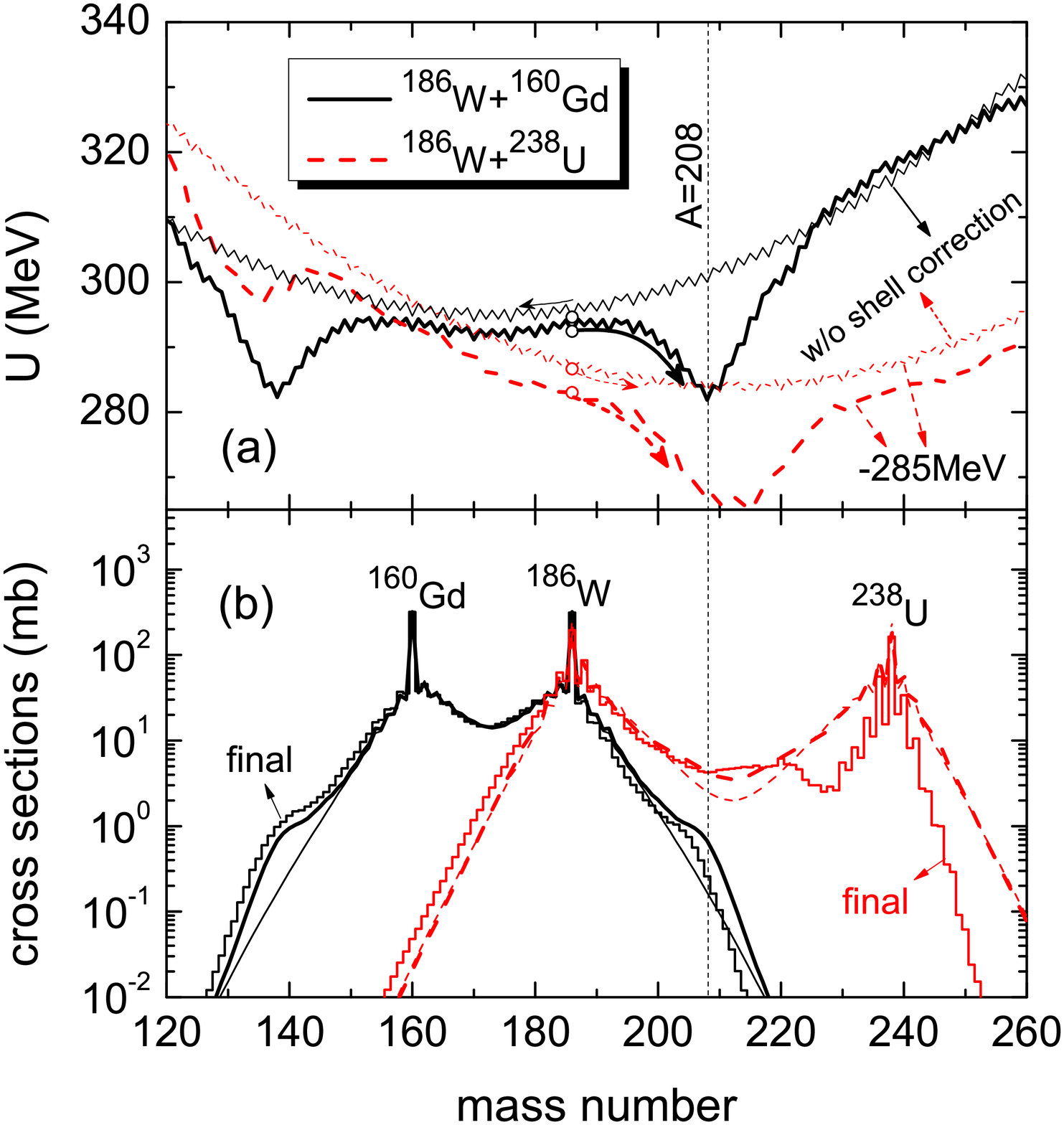}
\caption{\label{pot} (Color online.) (a) Potential energies as a function of mass number for the reactions $^{186}$W + $^{160}$Gd and $^{186}$W + $^{238}$U. The thick and thin lines denote the PES with and without shell correction, respectively. (b) Calculated mass distributions based on the DNS+GEMINI model in the reactions $^{186}$W + $^{160}$Gd and $^{186}$W + $^{238}$U at $E_{\textrm{c.m.}}=475$ and 660 MeV, respectively. The thick and thin lines denote mass distributions of primary fragments with and without shell correction, respectively. The solid histograms denote the mass distributions of final fragments with shell correction.}
\end{center}
\end{figure}

\begin{figure}
\begin{center}
\includegraphics[width=8.8cm,angle=0]{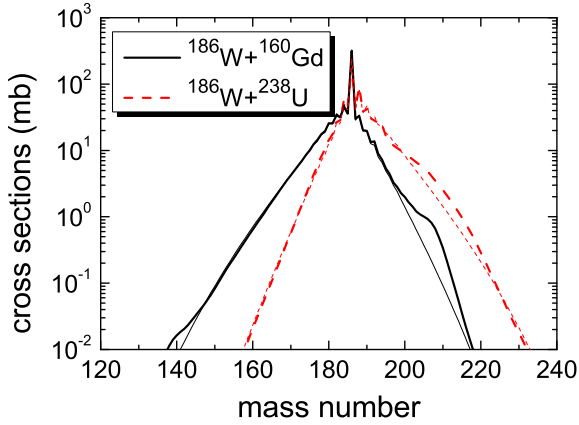}
\caption{\label{dis} (Color online.) The yields of primary fragments without contribution of TLF. The thick and thin lines denote the results of with and without shell correction, respectively.}
\end{center}
\end{figure}

The nucleon transfer process could be explained on the basis of PES. In Fig. \ref{pot}(a), the potential energy at the contact configuration is shown as a function of mass number for the reaction $^{186}$W + $^{160}$Gd. Due to shell closures of $N=126$, $N=82$, and $Z=82$, one deep pocket can be seen around the configuration of $^{208}$Pb + $^{138}$Ba. Due to attraction of this deep pocket, the inverse QF process is strongly promoted. The potential energy without shell correction is also shown in the reaction $^{186}$W + $^{160}$Gd. It can be seen that the minimum potential energy locates at the symmetry configuration, which means mass asymmetry relaxation could influence the nucleon flow from heavy fragment to lighter one in MNT process.
We also show the potential energies as a function of mass number in the reaction $^{186}$W + $^{238}$U. In order to conveniently compare with the reaction $^{186}$W + $^{160}$Gd, the curves of the potential energies in the reaction $^{186}$W + $^{238}$U are moved down by 285 MeV. One can see that both mass asymmetry relaxation and shell effects could enhance the probability of nucleons transferring from $^{238}$U to $^{186}$W. Therefore, we expect that the production cross sections of transprojectile nuclei would be strongly enhanced in MNT reactions with $^{238}$U target.

The mass distributions of primary fragments calculated within DNS+GEMINI model in the reaction $^{186}$W + $^{160}$Gd can be seen in Fig. \ref{pot}(b). Due to shell effects, one pronounced shoulder around $A=208$ is noticed. The same behavior was shown in Ref. \cite{Zagrebaev01}. After cooling process, the shoulder is still clear, only moves to smaller mass position. We also show the results based on $^{238}$U target for comparison. As we expected, the cross sections for producing transprojectile nuclei are much higher than those in the reaction $^{186}$W + $^{160}$Gd. In transprojectile region, for the curve without shell correction in the reaction $^{186}$W + $^{238}$U is still higher than that with shell enhancement in the reaction $^{186}$W + $^{160}$Gd. However, we cannot conclude that the mass asymmetry relaxation plays an significant role in nucleon flow so far. Actually, each mass distribution in Fig. \ref{pot}(b) contains TLF and PLF contributions. In the reaction $^{186}$W + $^{160}$Gd, the contribution in transprojectile region from TLF is very low. However, for the reaction $^{186}$W + $^{238}$U, because of mass asymmetry relaxation the yield of TLF around $A=210$ is almost half of the total.

Experimentally, it is difficult to detect and separate TLF in MNT reactions with present equipments. In order to clarify the mass asymmetry relaxation effect in MNT reactions, we remove the yield contribution of TLF and show in Fig. \ref{dis}. Clearly, the reaction $^{186}$W + $^{238}$U still shows larger cross sections in transprojectile region. Therefore, mass asymmetry relaxation significantly influences the nucleon flow in multinucleon transfer process. Also, the results of without shell correction are shown. The same behavior is notice as shown in Fig. \ref{pot}(b).
\begin{figure}[thb!]
\centering
\includegraphics[width=8.5cm,angle=0]{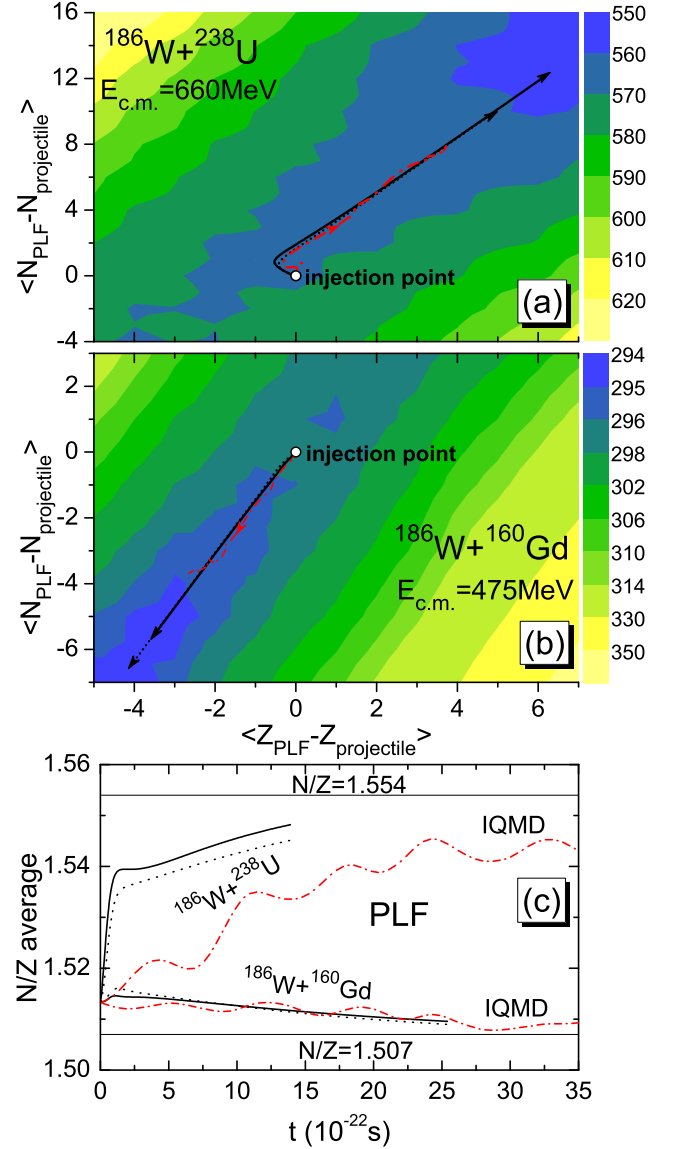}
\caption{\label{traj} (Color online.) Contour plots of the PES (in MeV) with the drift trajectories of the first moments of PLF distributions in a $\langle Z_{\textrm{PLF}}-Z_{\textrm{projectile}}\rangle$, $\langle N_{\textrm{PLF}}-N_{\textrm{projectile}}\rangle$ plane for the reactions $^{186}$W + $^{238}$U (a) and $^{186}$W + $^{160}$Gd (b). (c) Calculated average N/Z values of PLF as a function of relaxation time; For the IQMD model calculation, the relaxation process is initiated at contact configuration, which are about 90 fm/c and 100 fm/c for the reactions $^{186}$W + $^{238}$U and $^{186}$W + $^{160}$Gd, respectively; The N/Z values of compound systems are also shown with two horizontal lines. The dotted and dash-dotted lines denote the results of the DNS model without shell corrections and the IQMD model, respectively. The calculations are in head-on collisions ($J=0$).}
\end{figure}

In Fig. \ref{traj}(a), we show the contour of PES with evolution trajectory of the first moments of PLF distributions in a $\langle Z_{\textrm{PLF}}-Z_{\textrm{projectile}}\rangle$, $\langle N_{\textrm{PLF}}-N_{\textrm{projectile}}\rangle$ plane for the reaction $^{186}$W + $^{238}$U. The solid line denotes the result of the DNS model calculation. The tendency to minimize the potential energy is recognized. It can be seen that the target tends to lose nucleons to the projectile. However, the average neutron and proton numbers of PLF decreases along the trajectory in the reaction $^{186}$W + $^{160}$Gd, although shell effects promote the transferring of nucleons from target to projectile, as shown in Fig. \ref{traj}(b). The trajectories of the DNS model calculations without shell correction are also shown for both reactions with dotted lines. It can be seen the shell effects enhance the probability of transferring nucleons from target to projectile. However, mass asymmetry relaxation shows stronger effects on nucleon flow than the shell closures. For comparison, the trajectories of the IQMD model calculations are also shown in both reactions. For head-on collisions in IQMD model, the separation plane at a given time t is defined at the position where the isocontours of projectile and target densities cross \cite{Washiyama02}. No structure effect is considered in present IQMD model calculation. Nevertheless, the behavior of mass asymmetry relaxation also can be seen. The yields of transprojectile nuclei would be enhanced in $^{186}$W + $^{238}$U reaction. The dynamical simulations confirm the importance of mass asymmetry relaxation process in multinucleon transfer process.

As stated above, due to strong influence of mass asymmetry relaxation, large yields of transprojectile nuclei can be seen in the reaction $^{186}$W + $^{238}$U. For producing neutron-rich nuclei, one may wonder whether the neutron richness of PLF is high. To clarify this, we show the average $N/Z$ ratios of PLF in both reactions as a function of interaction time within the frameworks of the DNS model and the IQMD model in Fig. \ref{traj}(c). For the reaction $^{186}$W + $^{238}$U, the calculated average value of $N/Z$ ratio in the DNS model first increases quickly to the value of about 1.54, and then gently increases to 1.55, which is close to the $N/Z$ ratio of the compound system. It was found that the $N/Z$ equilibration occurs at the first stage of heavy ion collisions \cite{Gatty01,Hofmann01}. The DNS calculation approves the behavior of fast equilibration of $N/Z$ in heavy ion collisions, which is governed by gradient of PES. Also, it can be seen that shell effects significantly enhance the average $N/Z$ ratio of PLF in the reaction $^{186}$W + $^{238}$U, which is because of the shell closure $N=126$. Unlike strong increase of average $N/Z$ value of PLF within short relaxation time in the DNS model calculation, the $N/Z$ equilibration process evolves gradually in the IQMD model simulations. The fluctuation is mainly due to pre-equilibrium dipole oscillations \cite{Wu01}, which actually is supposed as the cause of $N/Z$ equilibration based on microscopic framework \cite{Bonche01}. On the other hand, due to charge equilibration, $^{238}$U ($N/Z=1.587$) enhances the neutron richness of PLF. For the reaction $^{186}$W + $^{160}$Gd, the $N/Z$ values of $^{160}$Gd and $^{186}$W are very close, which results in the almost flat variation of average $N/Z$ ratio with the interaction time in both the DNS and IQMD models calculations. Here, we would like to emphasize that in the IQMD model calculations, the relaxation processes are initiated at the contact configurations, which are 90 fm/c and 100 fm/c for the reactions $^{186}$W + $^{238}$U and $^{186}$W + $^{160}$Gd, respectively, after the beginning of simulations.
\begin{figure}[t]
\begin{center}
\includegraphics[width=8.5cm,angle=0]{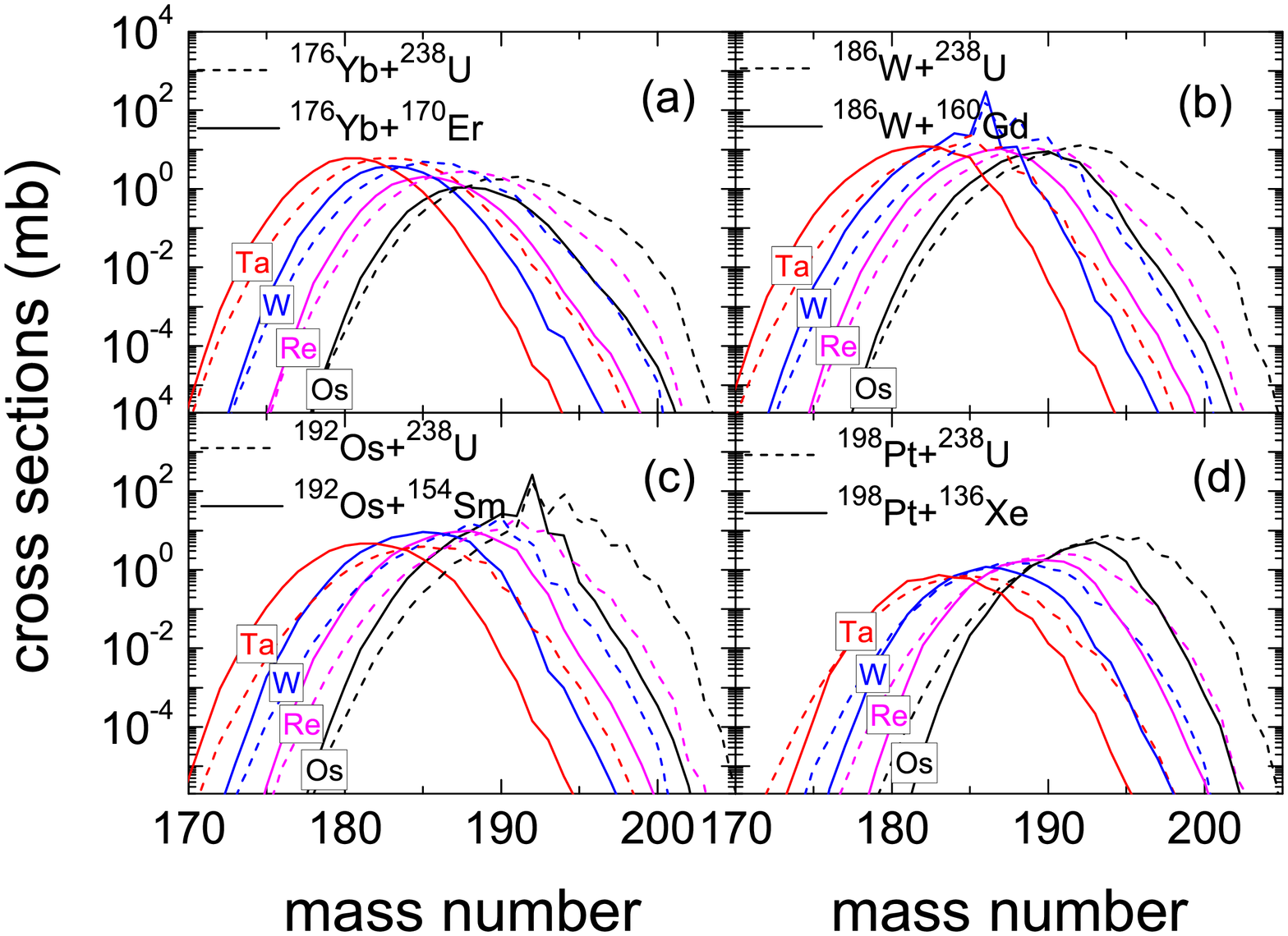}
\caption{\label{cross} (Color online.) Calculated production cross sections of Ta, W, Re, and Os isotopes in the reactions based on $^{176}$Yb (a), $^{186}$W (b), $^{192}$Os (c), and $^{198}$Pt (d) projectiles in the DNS+GEMINI model. }
\end{center}
\end{figure}

\begin{figure}[t]
\begin{center}
\includegraphics[width=8.8cm,angle=0]{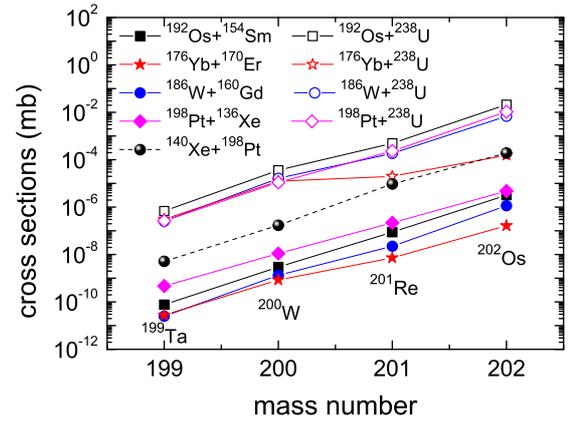}
\caption{\label{126} (Color online.) Calculated production cross sections of $^{202}$Os, $^{201}$Re, $^{200}$W, and $^{199}$Ta in the DNS+GEMINI model.}
\end{center}
\end{figure}

In Fig. \ref{cross}, we show the calculated production cross sections of Ta, W, Re, and Os isotopes in the reactions $^{176}$Yb + $^{238}$U, $^{176}$Yb + $^{170}$Er, $^{186}$W + $^{238}$U, $^{186}$W + $^{160}$Gd, $^{192}$Os + $^{238}$U, $^{192}$Os + $^{154}$Sm, $^{198}$Pt + $^{238}$U, and  $^{198}$Pt + $^{136}$Xe within the framework of the DNS+GEMINI model. As we expected, the reactions based on the $^{238}$U target show great advantages of cross sections for producing neutron-rich nuclei. We extract the production cross sections of nuclei with the neutron closed shell $N=126$ and show in Fig. \ref{126}. For producing $^{202}$Os, $^{201}$Re, $^{200}$W, and $^{199}$Ta, it can be seen that the production cross sections in the reactions based on the $^{238}$U target are almost 2 to 4 orders of magnitude larger than other reactions. We also show the calculated cross sections in the radioactive beam $^{140}$Xe induced reaction $^{140}$Xe + $^{198}$Pt at $E_{\textrm{c.m.}}=446.8$ MeV. Although $N/Z$ value of $^{140}$Xe is close to that of $^{238}$U, the production cross sections of isotopes $^{202}$Os, $^{201}$Re, $^{200}$W, and $^{199}$Ta are much lower than those in most of the reactions based on $^{238}$U target, except that the reaction $^{176}$Yb + $^{238}$U only shows comparable cross section with the reaction $^{140}$Xe + $^{198}$Pt for producing $^{202}$Os. This is because for producing $^{202}$Os only 2 protons transferring is needed in the reaction $^{140}$Xe + $^{198}$Pt, while 6 protons need to be transferred from target to the projectile in the system $^{176}$Yb + $^{238}$U.

\section{\label{summary}Summary}
For successfully obtaining the new neutron-rich isotopes around $N=126$, the MNT reactions based on $^{238}$U are investigated and compared with several promising reactions within the frameworks of the DNS (semi-classical approach) and IQMD (microscopic dynamical approach) models. The comparisons of DNS and IQMD models for descriptions of mass asymmetry relaxation and N/Z equilibration are given for the first time. Three conjectures about reactions with $^{238}$U are testified. (i) The mass asymmetry relaxation plays an significant role on transferring nucleons from $^{238}$U to light partners; (ii) The neutron closed shell $N=126$ could attract the neutrons flow from $^{238}$U to light partners; (iii) The $^{238}$U shows large value of $N/Z$ ratio and enhances the neutron-richness of projectile-like products. For the first time, the advantages for producing NRHN around $N=126$ in the MNT reactions with $^{238}$U are found and the MNT reactions with $^{238}$U target are proposed for producing unknown NRHN around $N=126$ in consideration of direct kinematics.

\textbf{Acknowledgments}
This work was supported by the National Natural Science Foundation of China under Grants No. 11605296; the Natural Science Foundation of Guangdong Province, China (Grant No. 2016A030310208); the National Natural Science Foundation of China under Grants No. 11805015, No. 11875328, No. 11605270, and No. 11805289.

%\begin{center}
%\section*{Figure Captions}
%\end{center}

%\begin{figure}
%\includegraphics[width=10cm,angle=0]{fig1.eps}
%\caption{\label{EOS}
%}
%\end{figure}

\end{document}